\documentclass[preprint]{aastex631}

\usepackage{amsmath}
\DeclareMathOperator*{\argmax}{argmax}
\DeclareMathOperator*{\argmin}{argmin}
\allowdisplaybreaks
\usepackage{xcolor}

\DeclareMathOperator{\arctantwo}{arctan2}

\submitjournal{ApJ}

\shorttitle{Analytical Inversion of Exoplanet Transit Spectra}
\shortauthors{Matcheva et al.}
\graphicspath{{./}{figures/}}
\begin{document}

\title{Transverse Vector Decomposition Method for Analytical Inversion of Exoplanet Transit Spectra}

\correspondingauthor{Katia Matcheva}
\email{matcheva@ufl.edu}

\author[0000-0003-4182-9096]{Konstantin T. Matchev}
\affiliation{Physics Department, University of Florida, Gainesville, FL 32653, USA}

\author[0000-0003-3074-998X]{Katia Matcheva}
\affiliation{Physics Department, University of Florida, Gainesville, FL 32653, USA}

\author[0000-0003-2719-221X]{Alexander Roman}
\affiliation{Physics Department, University of Florida, Gainesville, FL 32653, USA}



\begin{abstract}

We develop a new method for analytical inversion of binned exoplanet transit spectra and for retrieval of planet parameters. The method has a geometrical interpretation and treats each observed spectrum as a single vector $\vec r$ in the multidimensional spectral space of observed bin values. We decompose the observed $\vec r$ into a wavelength-independent component $\vec r_\parallel$ corresponding to the spectral mean across all observed bins, and a transverse component $\vec r_\perp$ which is wavelength-dependent and contains the relevant information about the atmospheric chemistry. The method allows us to extract, without {\em any} prior assumptions or additional information, the relative mass (or volume) mixing ratios of the absorbers in the atmosphere, the scale height to stellar radius ratio, $H/R_S$, and the atmospheric temperature. The method is illustrated and validated with several examples of increasing complexity.

\end{abstract}



%\tableofcontents

%\newpage % this is just to fix a (hopefully) temporary annoyance where only the header was on its own page!

\section{Introduction}

 Two and a half decades after the discovery of the first extrasolar system planet, the number of known exoplanets is now into the thousands\footnote{NASA Exoplanet Archive, \url{https://exoplanetarchive.ipac.caltech.edu/}.}. After the initial boom of discoveries based on the radial velocity method \citep{Wright2018,Perryman2018}, the current detections are mostly done using transit observations, where a planet is identified while passing in front of the host star, blocking a small fraction of the stellar flux \citep{Henry2000,Charbonneau2000}. Subsequently, some of the transiting planets are targeted for spectroscopic studies and the transit is recorded at different wavelengths \citep{Schneider1994,Seager2000}. A number of current and planned large-scale planetary surveys are based on observing {\em planetary transits} at different wavelengths, thus providing spectral information about the composition and the structure of the atmosphere of the planet \citep{Madhusudhan2019}.   

During a transit event, a planet blocks a certain fraction, $M(\lambda)$,  of the original stellar flux $F_O(\lambda)$,
\begin{equation}
M(\lambda) \equiv \frac{F_O(\lambda) - F_T(\lambda)}{F_O(\lambda)} = \left(\frac{R_T(\lambda)}{R_S}\right)^2,
\label{eq:Mdef}
\end{equation}
where $F_T(\lambda)$ is (the minimum of) the observed flux during transit at a given wavelength $\lambda$, $R_S$ is the stellar radius, and the transit radius $R_T(\lambda)$ is the radius of the equivalent solid disk blocking the same amount of light. The key feature of (\ref{eq:Mdef}) is the dependence of the transit depth $M(\lambda)$ on the observed wavelength $\lambda$, reflecting the differential absorption properties of the atmosphere due to its chemical composition. This allows for detection of atmospheric gasses with strong absorption lines in the probed wavelength range. In addition, the transit spectrum also contains information about the physical characteristics of the planet like its size, temperature, gravity, atmospheric pressure, etc.

The primary goal of exoplanet transit spectroscopy is the inversion of the observed spectrum in order to retrieve the parameters of the planet and its atmosphere \citep{Fisher2018MNRAS,Cobb2019,Barstow2020,Kitzmann2020,Harrington2021,Cubillos2021,Blecic2021,Welbanks2021,Welbanks2021atmospheric}. This inversion process inevitably relies on a forward model (FM), which, for a given set $\mathcal{S}$ of planet parameters, can predict the observed transit radius at each wavelength:
\begin{equation}
    \mathcal{S}(\textrm{planet parameters})
    \overset{FM}{\longrightarrow} R_T(\lambda).
\label{eq:forwardmodel}    
\end{equation}

The inverse problem corresponding to the forward model (\ref{eq:forwardmodel}) can then be formulated as follows: given the measured spectrum $R_T(\lambda)$, derive the values of the planet-specific parameters. This is a difficult challenge due to intrinsic theoretical complications: first of all, the problem is highly non-linear and does not easily lend itself to simple inversion methods from linear algebra; furthermore, there exist several directions of degenerate solutions in the parameter space which lead to equally plausible transit spectra. In this paper we shall focus on overcoming these theoretical challenges, leaving aside the additional observational complications due to instrumental noise, spectral resolution, etc.

The standard approach is to solve the inversion problem numerically. Classical inversion methods extensively used in analyzing infrared spectra from remote sensing of the Earth and Solar System planets are described in \cite{Hanel2003} and are based on linearizing the problem. More recent approaches incorporate various machine learning (ML) techniques
\citep{Marquez2018,Zingales2018,Cobb2019,Oreshenko2020,Fisher2020,Himes2020, Himes2020proc,Guzman2020,Nixon2020,Himes2020Marge,Yip2021,Ardevol2022} (for a recent comparative review, see \cite{Barstow2020MNRAS}). The main disadvantages of the numerical ML approach are: i) the ML model is in principle a black box which hides the relevant physics and is not easily interpretable \citep{Nixon2020,Yip2021}; ii) the ML model does not learn directly the relevant physics, but only the (finite amount of) data generated by the forward model, which introduces additional uncertainties during the training process \citep{Matchev:2020tbw}.

On the other hand, less has been done in terms of analytical approaches to the inverse problem. The first step in this direction would be to replace the numerical forward model (\ref{eq:forwardmodel}) with analytical expressions which capture the relevant physics \citep{Brown2001,Hubbard2001,Burrows2003,Fortney2005,Benneke2012,deWit2013,Griffith2014,Vahidinia2014,Heng2015,Betremieux2017,Heng2017,Heng2019MNRAS} and can then be possibly inverted. The analytical formulas typically rely on simplifying assumptions and approximations,.e.g, spherical symmetry, isothermal atmosphere, isobaric approximation, constant specific gravity, etc. Nevertheless, these theoretical results agree well with the numerical forward models and allow for better understanding of the underlying physics and of the intrinsic parameter degeneracies present in the problem \citep{Griffith2014,Heng2017,Welbanks2019,matchev2021analytical}.

The main goal of this paper is to push the limits of the analytical approach by applying ideas from vector algebra and analytical geometry to disentangle the parameter dependencies within the forward model and provide a clean measurement of the relative chemical abundances of the absorbers present in the atmosphere. This allows for direct and meaningful comparisons of the atmospheric {\bf chemical} compositions of different exoplanets, independent of the individual {\bf physical} planet parameters.

The paper is organized as follows. In section~\ref{sec:forward} we describe a typical forward model and introduce the relevant parameters. In section~\ref{sec:analytical} we review some of the existing analytical approximations in the literature, highlighting their generic features which will be relevant for our discussion. In section~\ref{sec:inversion} we describe our analytical inversion procedure which results in the measurement of the relative chemical abundances and two combinations of planet-specific physical parameters. In sections~\ref{sec:toy32}, \ref{sec:toy43}, \ref{sec:toy33} and \ref{sec:toy13} we illustrate our method with several toy examples of increasing complexity. In section \ref{sec:summary} we present our summary and conclusions.

\section{Forward Model}
\label{sec:forward}

The planet-specific input parameters to the forward model can be categorized into the following three groups:
\begin{itemize}
    \item {\em Physical parameters.} These are parameters which refer to the physical characteristics of the planet or the observational geometry: radius $R_S$ of the host star, atmospheric temperature $T$, specific gravity $g$ and reference pressure level $P_0$ at a given planet radius $R_0$ corresponding to an optically thick atmosphere along the line of sight \citep{Heng2017}.
    \item {\em Chemical composition parameters.} These are parameters which refer to the chemical composition of the planet atmosphere. The mean molecular mass $\bar m$ is determined by the major gas components (e.g., hydrogen and helium for hot Jupiters). The attenuation of the stellar light depends on the volume mixing ratio $X_a=n_a/n$ (or equivalently, the number density $n_a$) and the molecular mass $m_a$ of each individual absorber, where $a=1,2,\ldots,N_{abs}$ and $n$ is the total atmospheric number density.  
    \item {\em Wavelength-dependent parameters.} These include the cloud opacity $\kappa_{cl}(\lambda)$ and the absorption coefficients $\chi_a(\lambda)$ for the individual absorbers. In principle, these parameters can be viewed as a subset of the chemical composition parameters, but here we would like to emphasize their wavelength dependence. Note that the absorption coefficients $\chi_a$ in principle depend not only on the wavelength, but also on temperature and (to a lesser extent) on pressure.
\end{itemize}

With those preliminaries, we can now express the forward model (\ref{eq:forwardmodel}) more concretely as
\begin{equation}
    \mathcal{S}(R_S,R_0,T,P_0,g;\bar{m},X_a,m_a;\chi_a(\lambda),\kappa_{cl}(\lambda))\longrightarrow R_T(\lambda).
\label{eq:forwardmodel2}    
\end{equation}
Note that not all of the listed inputs are {\em a priori} unknown. For example, for a given constituent $a$, its molecular mass $m_a$ and absorption coefficients $\chi_a(\lambda)$ are known physical constants. In addition, alternative astronomical observations may provide independent constraints on some of the other input parameters. However, to be completely general, in this paper we shall not rely on such additional observations and will instead treat all planet-specific input parameters (i.e., everything except for the known physical constants $m_a$ and $\chi_a(\lambda)$) as 
{\em a priori} unknown.

The general expression (\ref{eq:forwardmodel2}) for the FM can be further simplified by noting that the absorption of the individual gases enters the model only through the total atmospheric opacity 
\begin{equation}
\kappa(\lambda) \equiv \kappa_{cl}(\lambda) + \sum_{a=1}^{N_{abs}} X_a \frac{m_a}{\bar m} \chi_a(\lambda),    
\label{eq:kappavec}
\end{equation}
where $N_{abs}$ is the number of absorbers considered in the model. Without any loss of generality, the forward model (\ref{eq:forwardmodel2}) can then be rewritten as
\begin{equation}
    \mathcal{S}\left(R_S,R_0,T,P_0,g;\bar{m},\kappa(\lambda)\right)\longrightarrow R_T(\lambda)
\label{eq:forwardmodel3}    
\end{equation}
in terms of the seven input parameters $R_S$, $R_0$, $T$, $P_0$, $g$, $\bar{m}$ and $\kappa(\lambda)$ \citep{matchev2021analytical}.

\section{Analytical Models}
\label{sec:analytical}

Equations~(\ref{eq:kappavec}) and (\ref{eq:forwardmodel3}) define the general setup for the inverse problem under consideration in this paper. In this section we narrow down the possible analytical forms of the forward model (\ref{eq:forwardmodel3}).

The first simplification arises due to standard dimensional analysis \citep{Langhaar1951,Barenblatt1996}. \cite{matchev2021analytical} showed that, again without any loss of generality, the analytical expression behind the forward model (\ref{eq:forwardmodel3}) must be of the form
\begin{equation}
    R_T(\lambda) = R_S \times
f\left(\frac{R_0}{H},\frac{P_0\, \kappa(\lambda)}{g},\frac{R_0}{R_S},\frac{ \bar{m}\, \kappa(\lambda)}{R_0^2}\right),
\label{eq:RTdiman}
\end{equation}
where $H$ is the pressure scale height 
\begin{equation}
H \equiv \frac{k_B T}{\bar{m} g},
\label{eq:Hdef}
\end{equation}
$k_B$ is the Boltzmann constant and $f$ is an unspecified function of dimensionless quantities only, i.e., of the four dimensionless Pi groups 
\begin{equation}
\pi_1 \equiv \frac{R_0}{H},
\qquad
\pi_2 \equiv \frac{P_0\, \kappa(\lambda)}{g},
\qquad
\pi_3 \equiv \frac{R_0}{R_S},
\qquad
\pi_4 \equiv \frac{ \bar{m} \kappa(\lambda)}{R_0^2}
\end{equation}
listed as its arguments, plus possibly some numerical dimensionless mathematical constants like $\pi$, $e$, etc. In order to provide the necessary length dimensions for $R_T$, in eq.~(\ref{eq:RTdiman}) we singled out $R_S$ as a prefactor; the same role could in principle be played by $H$ or $R_0$ (the latter choice was the one made in \cite{matchev2021analytical}). Here we find it convenient to work with $R_S$ because the transit depth $M$ is quoted in terms of the ratio $R_T/R_S$, see eq.~(\ref{eq:Mdef}).

\cite{matchev2021analytical} also showed that $\pi_4$ is not a relevant parameter in the problem and can be safely dropped. The end result from the dimensional analysis can therefore be written as
\begin{equation}
\sqrt{M(\lambda)} = \frac{R_T(\lambda)}{R_S} = f\left(\frac{R_0}{H},\frac{P_0\, \kappa(\lambda)}{g},\frac{R_0}{R_S}\right).
\label{eq:sqrtM}
\end{equation}

To proceed further, one has to perform explicit theoretical calculations which would reveal the specific form of the function $f$. In order to obtain relatively simple analytical results, one typically has to make some simplifying assumptions. In the case of an isothermal, radially symmetric atmosphere  \cite{Heng2017} obtained a simple analytical expression which can be written in analogy to (\ref{eq:sqrtM}) as
\begin{equation}
\frac{R_T(\lambda)}{R_S} = 
\frac{R_0}{R_S} \biggl\{1 + \frac{H}{R_0} 
\biggl[\gamma_E + E_1(\tau_0) 
%+ \ln \biggl(\frac{P_0\kappa(\lambda)}{g}\sqrt{2\pi \frac{R_0}{H}}\biggr)
+ \ln (\tau_0 )
\biggr]\biggr\},
\label{eq:fullformula}
\end{equation}
where $\gamma_E=0.577215665$ is the Euler–Mascheroni constant,
\begin{equation}
\tau_0(\lambda) \equiv \frac{P_0\kappa(\lambda)}{g}\sqrt{2\pi \frac{R_0}{H}}
\end{equation}
is the optical thickness of the atmosphere along the line of sight at the reference radius $R_0$ and
\begin{equation}
E_1(\tau_0) = \int_{\tau_0}^\infty \frac{e^{-t}}{t} dt
\end{equation}
is the exponential integral of the first order with argument $\tau_0$.

For an optically thick atmosphere (large $\tau_0$), the $E_1$ term is negligible
%$$\lim_{\tau_0\to \infty} E_1(\tau_0) = 0$$
and (\ref{eq:fullformula}) simplifies to
\begin{equation}
\frac{R_T(\lambda)}{R_S} = \frac{R_0}{R_S} \biggl\{1 + \frac{H}{R_0} 
\biggl[\gamma_E + \ln \biggl(\frac{P_0\kappa(\lambda)}{g}\sqrt{2\pi \frac{R_0}{H}}\biggr)
\biggr]\biggr\},
\label{eq:simpleformula}
\end{equation}
which is consistent with the general result (\ref{eq:sqrtM}). In the rest of the paper, we shall use the parametrization (\ref{eq:simpleformula}) for illustration of our inversion procedure.

\section{Inversion Procedure} 
\label{sec:inversion}

In general, a transit spectrum is taken at different wavelengths, binned in accordance with the spectral resolution of the instrument. Let $N_{bin}$ be the number of spectral bins whose central values form the wavelength vector array
\begin{equation}
\vec{\lambda} \equiv \left(\lambda_{1}, \lambda_{2}, \lambda_{3}, \ldots, \lambda_{N_{bin}}\right).
\label{eq:lambdavec}
\end{equation}
As pictured in this equation, in the remainder of this paper we shall be using vector notation to represent quantities with $N_{bin}$ components, i.e., variables whose values depend on the wavelength $\lambda$. For example, the vector array containing the $N_{bin}$ measured transit radii will be denoted as
\begin{equation}
    \vec{R}_T \equiv \left(R_T(\lambda_{1}), R_T(\lambda_{2}), R_T(\lambda_{3}), \ldots, R_T(\lambda_{N_{bin}})\right).
\label{eq:RTvec}    
\end{equation}
Since a transit event measures not $\vec R_T$ itself, but the ratio $\vec R_T/R_S$ (see eq.~(\ref{eq:Mdef})), we shall simplify the notation by defining
\begin{equation}
\vec r \equiv \frac{\vec R_T}{R_S}.
\label{eq:rdef}
\end{equation}
The dimensionless vector $\vec{r}$ contains the full information about a given transit observation. In our approach, any given transit spectrum is represented by a single point in the $N_{bin}$-dimensional spectral space and $\vec{r}$ is nothing but the position vector of this point, measured from the origin.

An important direction in the $N_{bin}$-dimensional spectral space is given by the universal vector
\begin{equation}
\vec{u} \equiv (\underbrace{1,1,1,\ldots, 1}_{N_{bin}}),
\end{equation}
which points in the direction of perfectly flat transit spectra which show no variation with wavelength. This case could correspond to a planet without any atmosphere, or to an atmosphere dominated by {\em gray} clouds, whose opacity is independent of wavelength and can therefore be written as 
\begin{equation}
\vec{\kappa}_{cl} \equiv \kappa_{gray}\, \vec{u}.
\end{equation}
The absorption coefficients $\chi_a(\lambda)$ of the individual absorbers are wavelength dependent and should therefore also be organized into a vector array. However, since absorption coefficients have physical dimensions, we would like to factor out a dimensionful constant $\chi$ which renders the remaining coefficients $\xi_a(\lambda)$ dimensionless:
\begin{equation}
\vec\chi_a \equiv \chi \, \vec \xi_a.
\label{eq:xidef}
\end{equation}
The actual numerical value of $\chi$ is not important and can be chosen for convenience, the simplest choice is simply $\chi = 1\ \frac{\rm m^2}{\rm kg}$ in SI units. With those conventions, the total opacity (\ref{eq:kappavec}) becomes
\begin{equation}
\vec{\kappa} = \kappa_{gray}\, \vec{u} + \sum_{a=1}^{N_{abs}} X_a \frac{m_a}{\bar m} \chi\, \vec{\xi}_a.  
\label{eq:kappavecgray}
\end{equation}

With the new vector array notation, the main result (\ref{eq:simpleformula}) for the transit spectrum can be rewritten as

\begin{equation}
\vec r %\equiv \frac{\vec{R}_T}{R_S} 
= \biggl\{\frac{R_0}{R_S} + \frac{H}{R_S} 
\biggl[\gamma_E + \ln \biggl(\sqrt{2\pi \frac{R_0}{H}}\biggr)
\biggr]\biggr\} \vec{u} + \frac{H}{R_S}\,
\left\{\ln \left(\frac{P_0}{g}\right)\vec{u} +
\overrightarrow{\ln (\vec{\kappa})}
\right\}
,
\label{eq:RTvectors}
\end{equation}
where in the last term the logarithm function acts component-wise on the vector array $\vec\kappa$ as follows
\begin{equation}
    \overrightarrow{\ln (\vec{\kappa})} \equiv (\ln \kappa_1, \ln \kappa_2, \ldots, \ln \kappa_{N_{bin}}),
\label{eq:ufunc}    
\end{equation}
producing a new vector array, as suggested by the long right arrow above the function name. This operation is identical to the way universal functions act on numpy arrays, but unfortunately, there is no established formal mathematical notation for it, and we shall use the long arrow notation depicted in eq.~(\ref{eq:ufunc}).

Let us now work out in detail the last term in eq.~(\ref{eq:RTvectors}). First, let us simplify the notation by incorporating the cloud opacity term into the sum in eq.~(\ref{eq:kappavecgray}). Defining
\begin{equation}
X_0 \equiv \frac{\kappa_{gray}}{\chi}; \qquad 
m_0 \equiv \bar m;  \qquad 
\vec{\xi}_0 \equiv \vec u,
\label{eq:clouds}
\end{equation}
we can treat the clouds as the $a=0$ absorber and rewrite (\ref{eq:kappavecgray}) in a more compact form
\begin{equation}
\vec{\kappa} = \sum_{a=0}^{N_{abs}} X_a \frac{m_a}{\bar m} \chi \vec{\xi}_a.  
\label{eq:kappavecall}
\end{equation}
Now substituting this into (\ref{eq:RTvectors}), we have
\begin{subequations}
\begin{eqnarray}
\vec{r}\, &=& \biggl\{\frac{R_0}{R_S} + \frac{H}{R_S} 
\biggl[\gamma_E + \ln \biggl(\sqrt{2\pi \frac{R_0}{H}}\biggr)
\biggr]\biggr\} \vec{u} + \frac{H}{R_S}\,
\left\{\ln \left(\frac{P_0\chi}{g}\right)\vec{u} +
\overrightarrow{\ln \left( \sum_{a=0}^{N_{abs}} X_a \frac{m_a}{\bar m} \vec{\xi}_a
\right)} \right\}\\
&=& \biggl\{\frac{R_0}{R_S} + \frac{H}{R_S} 
\biggl[\gamma_E + \ln \biggl(\sqrt{2\pi \frac{R_0}{H}}\biggr)
\biggr]\biggr\} \vec{u} + \frac{H}{R_S}\,
\left\{\ln \left(\mu_{abs}\frac{P_0\chi}{g}\right)\vec{u} +
\overrightarrow{\ln \left( \sum_{a=0}^{N_{abs}}  \frac{X_am_a}{\bar m\mu_{abs}} \vec{\xi}_a
\right)} \right\},~~~
\label{eq:RTvectorsb}
\end{eqnarray}
\end{subequations}
where to arrive at the second line we have multiplied and divided the argument of the logarithm in the last term with the quantity
\begin{equation}
\mu_{abs} \equiv \sum_{a=0}^{N_{abs}} X_a \frac{m_a}{\bar m} = \sum_{a=0}^{N_{abs}} \mu_a,
\label{eq:Xtotaldef}
\end{equation}
which is the total mass mixing ratio of all absorbers.

To further simplify the notation, let us introduce the {\em relative} mass mixing ratios $\omega_a$ for the different absorbers (which for brevity we shall simply refer to as ``weights"), which can be written in several equivalent ways: 
\begin{equation}
\omega_a \equiv \frac{\mu_a}{\mu_{abs}}
= \frac{\mu_a}{\sum_b^{N_{abs}} \mu_b}
= \frac{X_a m_a}{\sum_b^{N_{abs}} X_b m_b}
= \frac{n_a m_a}{\sum_b^{N_{abs}} n_b m_b}
= \frac{\rho_a}{\sum_b^{N_{abs}} \rho_b}, \quad a=0,1,2,\ldots,N_{abs},
\label{eq:xadef}
\end{equation}
where the mass mixing ratio $\mu_a$, the volume mixing ratio $X_a$, the mass density $\rho_a$ and the number density $n_a$ of each absorber are related by
\begin{equation}
    \mu_a = X_a \frac{m_a}{\bar m}= \frac{n_a m_a}{n \bar m} = \frac{\rho_a}{n \bar m},
\end{equation}
and $n$ is the atmospheric number density, so that $n \bar m$ is the (mass) density of the atmosphere. By definition, the weights $\omega_a$ are normalized to one:
\begin{equation}
\sum_{a=0}^{N_{abs}} \omega_a  = 1.
\label{eq:absnormalization}
\end{equation}
This allows us to rewrite (\ref{eq:RTvectorsb}) in the form
\begin{equation}
\vec r \, = \biggl\{ \frac{R_0}{R_S} + \frac{H}{R_S} 
\biggl[\gamma_E + \ln \biggl(\mu_{abs}\frac{P_0 \chi }{g}\sqrt{2\pi \frac{R_0}{H}}\biggr)
\biggr]\biggr\} \vec{u} + \frac{H}{R_S}\, 
\overrightarrow{\ln \left( \sum_{a=0}^{N_{abs}} \omega_a \vec{\xi}_a
\right)}.
\label{eq:RTvectordecomposition}
\end{equation}
Before proceeding with the last step of the analysis, let us take stock of what we have been able to accomplish with the last equation. The $N_{bin}$-dimensional vector $\vec r$ representing the measured transit spectrum has been decomposed into two components with the help of the dimensionless vectors $\vec u$ and $\vec\xi_a$. The first component in (\ref{eq:RTvectordecomposition}) is along the universal vector $\vec u$ and depends on {\em all} of the input parameters. The second component in (\ref{eq:RTvectordecomposition}) is constructed out of the known dimensionless vectors $\vec \xi_a$ and its direction is entirely determined by the dimensionless chemical composition weights $\omega_a$, while its magnitude in addition depends on the ratio $H/R_S$. Our main goal will be to {\em measure} these components from the data and thus determine the relative chemical composition $\omega_a$, the $H/R_S$ ratio and the combination of parameters specifying the component of $\vec r$ along $\vec u$.

Note that the two components in (\ref{eq:RTvectordecomposition}) are not orthogonal, since the vectors $\vec\xi_a$ are not necessarily orthogonal to $\vec u$ (in fact, $\vec \xi_0$ is $\vec u$ itself). Therefore, the next step is to decompose the (dimensionless) vector appearing in the last term of (\ref{eq:RTvectordecomposition})
\begin{equation}
\vec L (\omega_a)\equiv \overrightarrow{\ln \left( \sum_{a=0}^{N_{abs}} \omega_a \vec{\xi}_a
\right)}
\label{eq:Lvectordefinition}
\end{equation}
into a longitudinal component $\vec L_\parallel$ along $\vec u$ and a transverse component $\vec L_\perp$ orthogonal to $\vec u$:
\begin{equation}
\vec L (\omega_a) \equiv \vec L_\parallel (\omega_a) + \vec L _\perp (\omega_a) =  \underbrace{(\vec L \cdot \hat u)\, \hat u}_{\vec L_\parallel} + \underbrace{\vec L - (\vec L \cdot \hat u)\, \hat u }_{\vec L _\perp},  
\label{eq:Ldecomposition}
\end{equation}
where $\hat u \equiv \vec u / \sqrt{N_{bin}}$ is a unit vector along $\vec u$. Substituting (\ref{eq:Lvectordefinition}) and (\ref{eq:Ldecomposition}) into (\ref{eq:RTvectordecomposition}), we get
\begin{equation}
%\frac{\vec{R}_T}{R_S} 
\vec r = \biggl\{ \frac{R_0}{R_S} + \frac{H}{R_S} 
\biggl[\gamma_E 
+ \frac{\vec L\cdot \vec u}{N_{bin}}
+ \ln \biggl(\mu_{abs}\frac{P_0 \chi}{g}\sqrt{2\pi \frac{R_0}{H}}\biggr)
\biggr]\biggr\} \vec{u} + \frac{H}{R_S}\, \vec L _\perp (\omega_a).
\label{eq:RTvectorfinal}
\end{equation}
This is our main result. We are now ready to perform the desired measurements with the following simple algorithm, whose starting point is the observed spectrum $\vec r$:
\begin{enumerate}
\item {\em Compute the $\parallel$ component of $\vec r = \vec{R}_T/R_S$.} The first step is to project the measured transit spectrum vector $\vec r$ onto the universal direction $\vec u$. This results in 
\begin{equation}
    r_\parallel \equiv
    \vec r \cdot \hat u = \biggl\{ \frac{R_0}{R_S} + \frac{H}{R_S} \biggl[\gamma_E + \frac{\vec L\cdot \vec u}{N_{bin}} + \ln \biggl(\mu_{abs}\frac{P_0 \chi }{g}\sqrt{2\pi \frac{R_0}{H}}\biggr) \biggr]\biggr\} \sqrt{N_{bin}}.
\label{eq:RTparallelmeasurement}
\end{equation}    
The left-hand side is an experimentally observed quantity, which is equal to $\sqrt{N_{bin}}$ times the average of the spectrum across all wavelengths (this average is an informative quantity --- for example, in \cite{Matchev2022} it was shown to be the first principal component of synthetic datasets of transit spectra). The right-hand side of this equation is a theoretical quantity (i.e., function of input planet-specific parameters) which is the theoretical interpretation of this measurement. 
\item {\em Compute the $\perp$ component of $\vec r = \vec{R}_T/R_S$.} Next we obtain the component of $\vec r$ orthogonal to $\vec u$ by simply subtracting the mean:
\begin{equation}
\vec r_\perp \equiv 
\vec r - r_\parallel \hat u = \frac{H}{R_S}\, \vec L _\perp (\omega_a).
\label{eq:RTperpmeasurement}
\end{equation}    
Once again, the left-hand side is an experimentally observable quantity, in this case an $(N_{bin}-1)$-dimensional vector lying within the $(N_{bin}-1)$-dimensional hyperplane orthogonal to $\vec u$. Note that the right-hand side depends on relatively few input parameters, namely, the relative mass mixing ratios $\omega_a$ of the absorbers and on the ratio $H/R_S$. We shall now proceed to determine those parameters independently.
\item {\em Determine the relative mass mixing ratios $\omega_a$.} The key observation at this point is that the {\em direction} of the transverse vector $\vec L _\perp (\omega_a)$ is completely determined by the dimensionless weights $\omega_a$. At the same time, this direction is already known from the measurement (\ref{eq:RTperpmeasurement}) of $\vec r_\perp$. Therefore, {\em all} parameters $\omega_a$ can be simultaneously determined by requiring that the measured vector $\vec r_\perp$ is in the same direction as $\vec L _\perp (\omega_a)$. This can be accomplished, for example, by scanning over all possible trial values $\tilde \omega_a$ until the two vectors point in the same direction, i.e., until the relevant dot product is maximized:
\begin{equation}
\omega_a = \argmax_{\tilde \omega_a\in [0,1]} \left( \vec r_\perp  \cdot \left(\frac{\vec L _\perp (\tilde \omega_a)}{|\vec L _\perp (\tilde \omega_a)|} \right)\right).
\label{eq:abundances}
\end{equation}
As indicated in eq.~(\ref{eq:abundances}), in the rest of the paper we shall use a tilde to denote trial values for different parameters. On the other hand, retrieved parameter values will carry no tilde.
\item {\em Determine the ratio $H/R_S$.} Once the relative weights are found from (\ref{eq:abundances}), the vector $\vec L _\perp (\omega_a)$ is fully determined. Then the ratio $H/R_S$ can be obtained by simply dividing the magnitudes (or the corresponding individual components) of $\vec r_\perp$ and $\vec L _\perp (\omega_a)$:
\begin{equation}
\frac{H}{R_S} = \frac{\left| \vec r_\perp \right|}{|\vec L_\perp (\omega_a) |}.
\label{eq:HRSmeasurement}
\end{equation}
\end{enumerate}

To review the bidding so far, the above algorithm has allowed us to measure a total of $N_{abs}+1$ degrees of freedom, namely: i) the parameter combination in the right-hand side of eq.~(\ref{eq:RTparallelmeasurement}), ii) the ratio $H/R_S$ and iii) the $N_{abs}-1$ independent degrees of freedom contained in the relative mass mixing ratios $\omega_a$ (recall that one degree of freedom there is removed by the normalization condition (\ref{eq:absnormalization})). Therefore, a necessary condition for being able to fully determine all these degrees of freedom is that the number of bins in the transit spectrum is at least that many:
\begin{equation}
N_{bin} \ge N_{abs} + 1.    
\label{eq:Nbincondition}
\end{equation}
In the next few sections we shall illustrate the algorithm described above with some simple yet non-trivial examples: $N_{bin}=3$ and $N_{abs}=2$ in section~\ref{sec:toy32}, $N_{bin}=4$ and $N_{abs}=3$ in section~\ref{sec:toy43} and $N_{bin}=3$ and $N_{abs}=3$ in section~\ref{sec:toy33}. In those cases, our inversion procedure can be easily visualized due to the relatively low number of bins. Then in section~\ref{sec:toy13} we shall consider a more realistic example with $N_{bin}=13$ and $N_{abs}=3$. In all of these examples, we shall use common fixed values for the following planet-specific parameters: $R_0 = 1.79 R_J$, $R_S = 1.57 R_{\odot}$, $g=9.77$ m/s$^{2}$, $P_0=10$ bar, $\bar m = 2.4\, m_{\rm amu}$, where $m_{\rm amu}=1.660539040\times 10^{-27}$ kg is the atomic mass unit, which are consistent with those inferred for WASP-12b \citep{Marquez2018}. Then for each model, we shall be choosing a study point defined by the masses $m_a$, the volume mixing ratios $X_a$ and the dimensionless absorption coefficients $\vec\xi_a$ (at a given temperature $T$) for the respective number of absorbers $N_{abs}$ and number of wavelength bins $N_{bin}$. Then we shall apply our algorithm to recover the values of the input parameters for the respective study point.

\section{Example with Three Spectral Bins and Two Absorbers} 
\label{sec:toy32}

We start with a simple toy example with $N_{bin}=3$ bins in the transit spectrum and $N_{abs}=2$ absorbers in the atmosphere. The two absorbers will be denoted as $A$ and $B$ and will be taken to have the following parameters at $T=2000$ K: 
\begin{subequations}
\begin{eqnarray}
\vec\xi_A = (10^{-3},10^{-3},10^{-1}), \quad m_A &=& 10\, m_{amu}, \quad X_A = 10^{-2} ~\Longrightarrow~ \omega_A = \frac{5}{6}\approx 0.833;
\label{Ex1Ainputs}\\
\vec\xi_B = (10^{-4},10^{-2},10^{-3}), \quad m_B &=& 20\, m_{amu}, \quad X_B = 10^{-3} ~\Longrightarrow~ \omega_B =\frac{1}{6} \approx 0.167.
\label{Ex1Binputs}
\end{eqnarray}
\end{subequations}
% M = 1.45598534 1.47292279 1.52868486 with exact inputs
The resulting spectrum from eq.~(\ref{eq:simpleformula}) is 
\begin{equation}
M(\lambda_1) = 1.455985, \quad
M(\lambda_2) = 1.472923, \quad
M(\lambda_3) = 1.528685,
\end{equation}
which corresponds to a vector $\vec r$ in the 3-bin spectral space
\begin{equation}
% R_T/R_S = 0.12066422 0.12136403 0.12364
\vec r = (0.120664, 0.121364, 0.123640).
\label{eq:rstudyexample1}
\end{equation}
We now decompose this vector into a longitudinal and transverse components:
\begin{equation}
\vec r = \vec r_\parallel + \vec r_\perp \equiv r_\parallel\hat{u} + r_{\perp 1}\,\hat{e}_1 +  r_{\perp 2}\,\hat{e}_2,
\end{equation}
where the unit vectors $\hat{u}$, $\hat{e}_1$ and $\hat{e}_1$ form an orthonormal right-handed system. The unit vector $\hat u$ was already discussed earlier, while the unit vectors $\hat e_1$ and $\hat e_2$ parametrize the transverse plane and one possible choice is as follows
\begin{equation}
\hat u = \frac{1}{\sqrt{3}} (1,1,1), \qquad
\hat e_1 = \frac{1}{\sqrt{2}} (0,1,-1), \qquad
\hat e_2 = \frac{1}{\sqrt{6}} (-2,1,1).
\label{eq:unitvectors3}
\end{equation}
Dotting (\ref{eq:rstudyexample1}) into (\ref{eq:unitvectors3}), we find the three components of $\vec{r}$
% 0.21111866
% [-0.00160935,  0.00150055]
\begin{equation}
r_\parallel =  2.11119\times 10^{-1}, \qquad
r_{\perp 1} = -1.60935\times 10^{-3}, \qquad
r_{\perp 2} =  1.50055\times 10^{-3}.
\label{eq:rcomponentsExample1}
\end{equation}
Using the numerical values for the planet parameters listed at the end of section~\ref{sec:inversion}, one can easily verify that the obtained numerical value for $r_\parallel$ is precisely as predicted by the right-hand side of eq.~(\ref{eq:RTparallelmeasurement}). This completes Steps 1 and 2 of the algorithm.

\begin{figure*}[t]
\centering
\includegraphics[height=.4\textwidth]{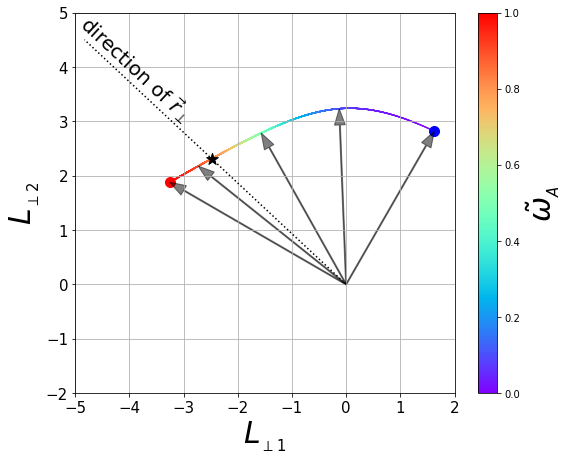}
\includegraphics[height=.4\textwidth]{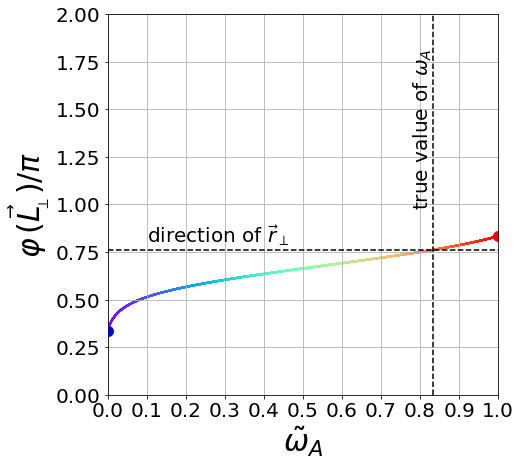}
\caption{Illustration of the extraction of the relative mass mixing ratios $w_A$ and $\omega_B=1-\omega_A$ and the ratio $H/R_S$ for the toy example with three spectral bins and two absorbers considered in section~\ref{sec:toy32}. The left panel depicts the family of vectors $\vec L_\perp(\tilde\omega_A, \tilde\omega_B)$, color-coded by the value of $\tilde\omega_A$. The dotted line marks the direction of $\vec r_\perp$ defined by (\ref{eq:phiEx1}). The right panel shows the relationship between the trial value $\tilde\omega_A$ and the resulting direction of $\vec L_\perp(\tilde\omega_A, \tilde\omega_B)$ in terms of its polar angle $\varphi$ (in units of $\pi$).
\label{fig:Ex1_Lperp}}
\end{figure*}

The measurement of the mixing ratios $\omega_A$ and $\omega_B$ via Step 3 is illustrated in the two panels of Figure~\ref{fig:Ex1_Lperp}. The direction of $\vec{r}_{\perp}$ in the transverse plane can be found from the measured components (\ref{eq:rcomponentsExample1}) and is given by the polar angle
% 0.7611311787250284
\begin{equation}
\varphi(\vec{r}_\perp) = \arctantwo \left(r_{\perp 2}, r_{\perp 1} \right) = 0.761 \pi.
\label{eq:phiEx1}
\end{equation}
This direction is indicated by a diagonal dotted line in the left panel of Figure~\ref{fig:Ex1_Lperp} and by a horizontal dashed line in the right panel of Figure~\ref{fig:Ex1_Lperp}. In order to find the weights $\omega_A$ and $\omega_B$, we need to build the set of vectors (\ref{eq:Lvectordefinition}) in terms of the trial values $\tilde\omega_A$ and $\tilde\omega_B$, i.e.
\begin{equation}
\vec L (\tilde\omega_A, \tilde\omega_B)\equiv \overrightarrow{\ln \left( \tilde\omega_A\, \vec{\xi}_A + \tilde\omega_B\, \vec{\xi}_B \right)},
\end{equation}
and find which one of them has a transverse component $\vec L_\perp(\tilde\omega_A, \tilde\omega_B)$ matching the direction of $\vec r_\perp$. This simple exercise is also illustrated in Figure~\ref{fig:Ex1_Lperp}. The colored points in the left panel of the figure represent the family of vectors $\vec L_\perp(\tilde\omega_A, \tilde\omega_B)$, color coded by the trial value of $\tilde\omega_A$ (recall that in this toy example $\tilde\omega_B$ will be simply given by $1-\tilde\omega_A$). The gray arrows depict five $\vec L_{\perp}$ vectors corresponding to five representative trial values of $\tilde\omega_A=\{1.0, 0.90, 0.50, 0.10, 0.0\}$ (from left to right), and the color coding gives all intermediate trial values of $\tilde \omega_A$. The direction of $\vec{r}_{\perp}$ intersects the $\vec L_{\perp}$ family at the point marked with the star symbol, which reveals the true value of $\omega_A$ as $\omega_A=0.833$, in precise agreement with the input value in (\ref{Ex1Ainputs}).

The right panel in Figure~\ref{fig:Ex1_Lperp} shows an alternative representation of the same measurement. Since the family of $\vec L_\perp(\tilde\omega_A, \tilde\omega_B)$ vectors does not involve any other planet-specific parameters, the relationship between the trial mass mixing ratio $\tilde\omega_A$ and the corresponding $\vec L_{\perp}(\tilde\omega_A, \tilde\omega_B)$ direction can be built up in advance, as shown in the right panel of Figure~\ref{fig:Ex1_Lperp}, where we kept the color-coding for convenience only (it simply tracks the horizontal axis). Then, once the direction of $\vec r_\perp$ is known from (\ref{eq:phiEx1}), we can directly read off the true value of the mixing ratio $\omega_A$ from the horizontal axis, as illustrated with the dashed lines.

Once the value of $\omega_A$ is known, in Step 4 of the algorithm we construct the vector
% [[-2.48092367], [ 2.31320635]]
\begin{equation}
\vec L_\perp(\omega_A=0.833) = -2.4809\, \hat e_1 + 2.3132\, \hat e_2,
\end{equation}
which allows us to compute the ratio $H/R_S$ in several equivalent ways:
\begin{equation}
\frac{H}{R_S} 
= \frac{r_{\perp 1}}{L_{\perp 1}} 
= \frac{r_{\perp 2}}{L_{\perp 2}}
= \frac{|\vec r_{\perp}|}{|\vec L_{\perp}|} = 0.0006487,
\end{equation}
which was precisely the input value in this example.

This concludes the demonstration of the algorithm in this very first toy example. The algorithm reproduced exactly (within machine precision) the input values for the fractional chemical abundance of the two components ($\omega_A$ and $\omega_B$), the ratio $H/R_S$ and the spectral average $r_\parallel$. The algorithm would work for any number of bins $N_{bin}$ and any number of chemical absorbers $N_{abs}$, yet the simplicity of this toy example allowed for easy visualization and comprehensive step-by-step demonstration of the main steps of the algorithm. In the following two sections we shall consider more complicated and more realistic examples, skipping some of the details already seen here and instead focusing on the new elements in those examples.

\section{Example with Four Spectral Bins and Three Absorbers} 
\label{sec:toy43}

In this section we shall consider a slightly more complex example with $N_{bin}=4$ bins in the transit spectrum and $N_{abs}=3$ absorbers in the atmosphere. The wavelength-dependent vector arrays now have 4 components. The three absorbers will be denoted as $A$, $B$ and $C$ and will be taken to have the following parameters (at $T=2000$ K)
% w_A= 0.15873015873015872 
% w_B= 0.31746031746031744 
% w_C= 0.5238095238095237
\begin{subequations}
\begin{eqnarray}
\vec\xi_A = (10^{-3},10^{-3},10^{-1},10^{-4}), \quad
m_A &=& 10\, m_{amu}, \quad X_A = 10^{-2} ~\Longrightarrow~ \omega_A = 0.159;\\
\vec\xi_B = (10^{-4},10^{-2},10^{-3},10^{-1}), \quad
m_B &=& 20\, m_{amu}, \quad X_B = 10^{-2} ~\Longrightarrow~ \omega_B = 0.317;\\
\vec\xi_C = (10^{-1},10^{-4},10^{-4},10^{-2}), \quad
m_C &=& 33\, m_{amu}, \quad X_C = 10^{-2} ~\Longrightarrow~ \omega_C = 0.524.
\end{eqnarray}
\label{eq:ex2inputs}
\end{subequations}
With those inputs, the transit 4-bin spectrum is 
% study point X_a = 0.01, 0.01, 0.01
% M = [1.54792266 1.50396958 1.52902229 1.5422581 ]
\begin{equation}
M(\lambda_1) = 1.547923, \quad
M(\lambda_2) = 1.503970, \quad
M(\lambda_3) = 1.529022, \quad
M(\lambda_4) = 1.542258,
\end{equation}
which corresponds to the vector $\vec r$ 
\begin{equation}
% new study point X_a = 0.01, 0.01, 0.01
% 0.12441554 0.12263644 0.12365364 0.12418768
\vec r = (0.1244155,\, 0.1226364,\, 0.1236536,\, 0.1241877), 
\label{eq:rstudyexample2}
\end{equation}
which we decompose as
\begin{equation}
\vec r = \vec r_\parallel + \vec r_\perp \equiv r_\parallel\hat{u} + r_{\perp 1}\,\hat{e}_1 +  r_{\perp 2}\,\hat{e}_2 + r_{\perp 3}\,\hat{e}_3
\end{equation}
onto the four orthonormal vectors
\begin{equation}
\hat u = \frac{1}{2}\, (1,1,1,1), \quad
\hat e_1 = \frac{1}{\sqrt{2}}\, (0,1,-1,0), \quad
\hat e_2 = \frac{1}{\sqrt{6}}\, (-2,1,1,0), \quad
\hat e_3 = \frac{1}{\sqrt{12}}\, (1,1,1,-3),
\label{eq:unitvectors4}
\end{equation}
resulting in the four components
% new study point X_a = 0.01, 0.01, 0.01
% r_parallel = [0.24744665]
% r_perp = [-0.00071927, -0.00103736, -0.0005362 ]
\begin{equation}
r_\parallel =  2.4745\times 10^{-1}, \quad
r_{\perp 1} = -7.1927\times 10^{-4}, \quad
r_{\perp 2} = -1.0374\times 10^{-3}, \quad
r_{\perp 3} = -5.3620\times 10^{-4}.
\label{eq:rcomponentsExample2}
\end{equation}
It is easy to check that the measured component $r_\parallel$ along the universal vector $\vec u$ once again exactly matches the prediction from the theoretical formula (\ref{eq:RTparallelmeasurement}). This time the transverse vector $\vec r_\perp$ is three-dimensional. In order to extract the chemical composition from its direction, we form the vector family
\begin{equation}
\vec L (\tilde\omega_a)\equiv \overrightarrow{\ln \left( \tilde\omega_A\, \vec{\xi}_A + \tilde\omega_B\, \vec{\xi}_B + \tilde\omega_C\, \vec{\xi}_C \right)}
\end{equation}
parametrized in terms of the trial values $\tilde\omega_A$, $\tilde\omega_B$ and $\tilde\omega_C=1-\tilde\omega_A-\tilde\omega_B$,
and project it on the transverse space as
\begin{equation}
\vec L_\perp (\tilde\omega_a) 
= \vec L(\tilde\omega_a) - (\vec L(\tilde\omega_a) \cdot \hat u)\, \hat u 
= \left( 
\vec L(\tilde\omega_a)\cdot \hat e_1,\, 
\vec L(\tilde\omega_a)\cdot \hat e_2,\,
\vec L(\tilde\omega_a)\cdot \hat e_3\right).
\label{eq:LperpsEx2}
\end{equation}

\begin{figure}[t]
\centering
\includegraphics[width=.49\textwidth]{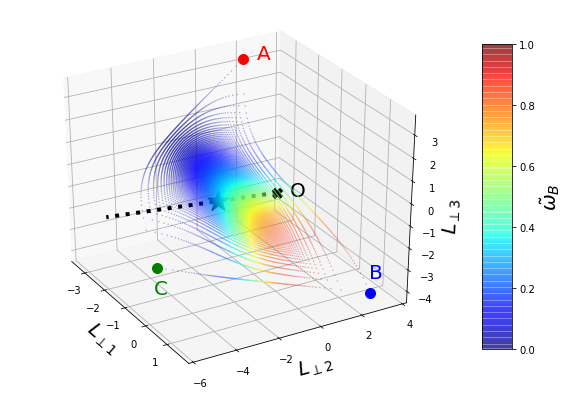}
\includegraphics[width=.49\textwidth]{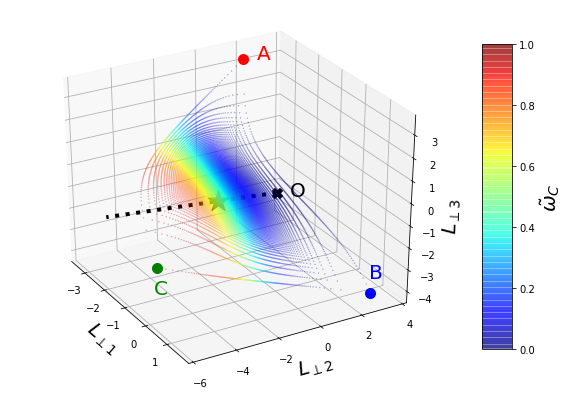}
\caption{The same as the left panel in Figure~\ref{fig:Ex1_Lperp}, but for the example discussed in Section~\ref{sec:toy43}. The three-dimensional family of transverse vectors (\ref{eq:LperpsEx2}) is color-coded by the trial value $\tilde\omega_B$ (left panel) or $\tilde\omega_C$ (right panel). The dotted line emerging from the origin {\bf O} of the coordinate system shows the measured direction of $\vec r_\perp$ according to (\ref{eq:rcomponentsExample2}).
\label{fig:LperpsEx2}}
\end{figure}

The family of transverse vectors $\vec L_\perp(\tilde\omega_a)$ is visualized in Figure~\ref{fig:LperpsEx2} in complete analogy with Figure~\ref{fig:Ex1_Lperp}. We scan $\tilde\omega_A$ and $\tilde\omega_B$ on a regular grid with 200 points in each direction and show the resulting grid of transverse vectors $\vec L_\perp (\tilde\omega_a)$. The tips of the vectors $\vec L_\perp (\tilde\omega_a)$ outline a two-dimensional triangularly-shaped surface in the three-dimensional transverse space, whose corners $A$, $B$ and $C$ correspond to $\tilde\omega_A=100\%$, $\tilde\omega_B=100\%$ and $\tilde\omega_C=100\%$, respectively. The surface is color-coded by the value of $\tilde \omega_B$ in the left panel and $\tilde \omega_C$ in the right panel. The origin {\bf O} of the coordinate system is marked with the $\times$ symbol and the dotted line originating from {\bf O} shows the measured direction of $\vec r_\perp$ according to (\ref{eq:rcomponentsExample2}). This line intersects the plotted surface at the point marked with the $\star$ symbol, which reveals the location of the vector $\vec L_\perp (\omega_a)$ built from the true values of the mass mixing ratios $\omega_a$. The length of this vector in turn reveals the ratio $H/R_S$ according to (\ref{eq:HRSmeasurement}). Once again, the described procedure resulted in {\em exact} retrieval of the parameters $\omega_a$, $H/R_S$ and $r_\parallel$ (the precision is only limited by the resolution of the $\tilde\omega_a$ scan which can be taken to be arbitrarily precise).

\begin{figure}[t]
\centering
\includegraphics[width=.6\textwidth]{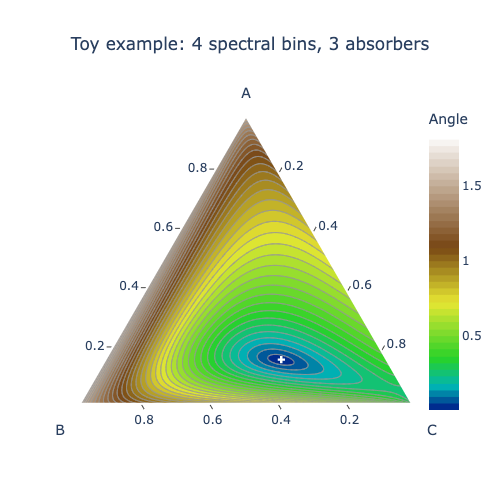}
\caption{A ternary plot illustrating the retrieval of the relative mass mixing ratios $\omega_A$, $\omega_B$ and $\omega_C$ for the example discussed in Section~\ref{sec:toy43}. Each point on the plot represents a unique combination of the trial values $\tilde\omega_A$, $\tilde\omega_B$ and $\tilde\omega_C$, subject to $\tilde\omega_A +\tilde\omega_B + \tilde\omega_C = 1$, with a scan resolution of $\Delta\, \tilde\omega=0.005$. The points are color-coded by the value of the relative angle $\Delta \Theta$ (in radians) between the fixed measured vector $\vec r_\perp$ and the trial vector $\vec L_\perp (\tilde\omega_A, \tilde\omega_B, \tilde\omega_C)$. The white cross marks the true values of $\omega_A$, $\omega_B$ and $\omega_C$. 
\label{fig:TernaryEx2}}
\end{figure}

Since in this example we have three absorbers with relative weights $\omega_a$ subject to $\omega_A+\omega_B+\omega_C=1$, the appropriate generalization of the right panel in Figure~\ref{fig:Ex1_Lperp} is a ternary plot, as illustrated in Figure~\ref{fig:TernaryEx2}. Each point on the plot represents a unique combination of the trial values $\tilde\omega_A$, $\tilde\omega_B$ and $\tilde\omega_C$, with the white cross denoting the true input values from (\ref{eq:ex2inputs}). These combinations are then color-coded by the value of the relative angle 
\begin{equation}
\Delta \Theta \equiv \arccos \left( \frac{\vec{r}_\perp}{|\vec{r}_\perp|}  \cdot \left(\frac{\vec L _\perp (\tilde \omega_a)}{|\vec L _\perp (\tilde \omega_a)|} \right)\right)
\label{eq:deltathetadef}
\end{equation}
between the fixed measured vector $\vec r_\perp$ and the trial vector $\vec L_\perp (\tilde\omega_A, \tilde\omega_B, \tilde\omega_C)$. The magnitude of $\Delta \Theta$ is indicative of the goodness of fit --- the smaller the value of $\Delta \Theta$, the more aligned the two vectors are.

Figure~\ref{fig:TernaryEx2} reveals that even a coarse scan of the relative mixing ratios with the current step of $\Delta \omega =0.005$ successfully finds the right answer. The best match is observed for $\tilde\omega_A=0.16$, $\tilde\omega_B=0.32$ and $\tilde\omega_C=0.52$, which is consistent with the inputs in (\ref{eq:ex2inputs}) within the scan resolution. Since our method is exact, in theory the precision of the retrieval can be arbitrarily improved by increasing the scan resolution. In practice, however, this will be limited by the observational noise level. 

\section{Underconstrained Example: Three Spectral Bins and Three Absorbers} 
\label{sec:toy33}

\begin{figure}[t]
\centering
\includegraphics[height=.43\textwidth]{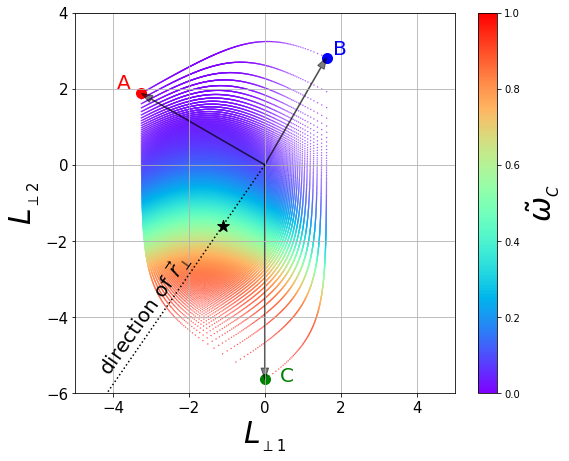}
\includegraphics[height=.43\textwidth]{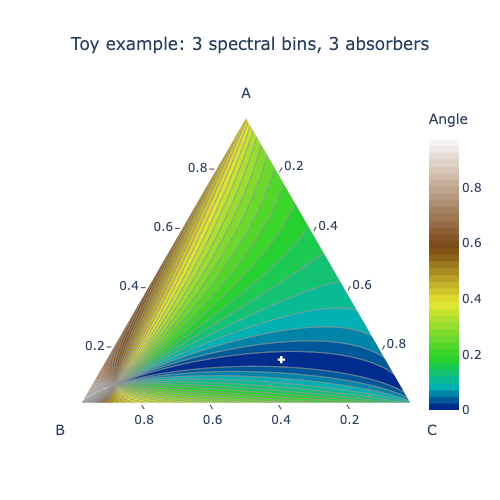}
\caption{Results for the underconstrained example discussed in Section~\ref{sec:toy33}. The left panel shows the trial family of vectors $\vec L_\perp(\tilde\omega_a)$ (scanned with resolution $\Delta\tilde\omega=0.002$), in complete analogy to the left panel of Figure~\ref{fig:Ex1_Lperp} and the right panel of Figure~\ref{fig:LperpsEx2}. Here the transverse space is two-dimensional (as in Figure~\ref{fig:Ex1_Lperp}) and the locus of points representing $\vec L_\perp(\tilde\omega_a)$ is two-dimensional as well (as in Figure~\ref{fig:LperpsEx2}). The measured direction of $\vec r_\perp$ indicated by the black dotted line is in the plane of the figure, and the $\star$ symbol marks the true value of $\vec L_\perp$. The right panel shows the corresponding ternary plot illustrating the results from the retrieval for this case, in analogy to Figure~\ref{fig:TernaryEx2}.
\label{fig:Ex2_degeneracy_ternary}}
\end{figure}

\begin{figure}[t]
\centering
\includegraphics[width=.47\textwidth]{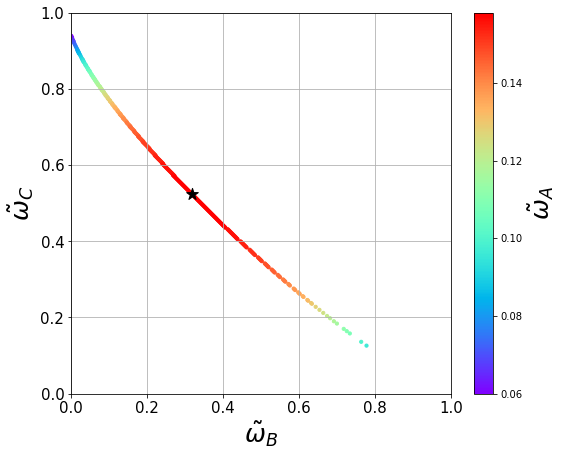}
\includegraphics[width=.47\textwidth]{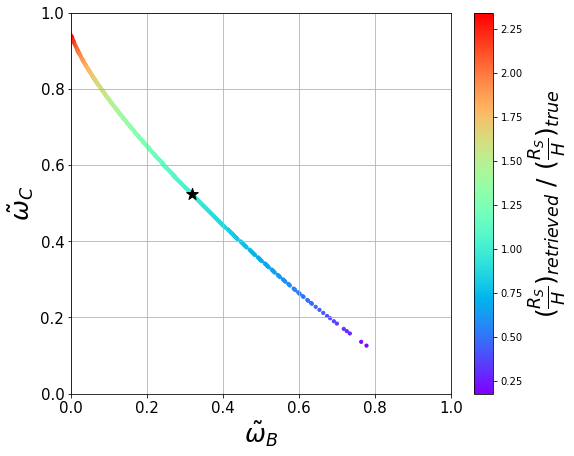}
\caption{Representation of the degenerate solutions found in the retrieval analysis of Section~\ref{sec:toy33}. All points from the $\vec L_\perp (\tilde \omega_a)$ grid with $\Delta\Theta < 0.001$ are plotted in the $(\tilde\omega_B, \tilde\omega_C)$ plane and color-coded by the value of $\tilde \omega_A$ (left panel) or the ratio of ratios (\ref{eq:ratioofratios}) (right panel).
\label{fig:Ex2_degeneracy}}
\end{figure}

The retrievals in the examples from the previous two sections were successful because we had enough spectral information and the condition (\ref{eq:Nbincondition}) was satisfied. Let us now briefly investigate how our method performs when this condition is violated and the system is underconstrained. For this purpose, let us reconsider the example from the previous section ($N_{bin}=4$ and $N_{abs}=3$), only this time let us omit the last spectral bin from the discussion, leaving us with only three bins ($N_{bin}=3$). The analysis proceeds the same way as before, only now the transverse space is two dimensional, as illustrated in the left panel of Figure~\ref{fig:Ex2_degeneracy_ternary}, which is the analogue of the plots in the left panel of Figure~\ref{fig:Ex1_Lperp} and in the right panel of Figure~\ref{fig:LperpsEx2}. Unlike those cases, however, this time the measured direction of the transverse vector $\vec r_\perp$ does not pick out a single solution for $\vec L_\perp(\tilde\omega_A,\tilde\omega_B,\tilde\omega_C)$. Instead, there is a whole family of vectors $\vec L_\perp(\tilde\omega_A,\tilde\omega_B,\tilde\omega_C)$ which are perfectly aligned with $\vec r_\perp$. This is also visible in the ternary plot in the right panel of Figure~\ref{fig:Ex2_degeneracy_ternary} (the analogue of Figure~\ref{fig:TernaryEx2}), which exhibits a whole region of possible solutions --- see the dark blue shaded valley with $\Delta\Theta=0$. Of course, the true solution, marked with the white cross, is also included in that region, but is not uniquely determined. Despite the degeneracy of the found solutions, the method is still capable of ruling out large areas of the ternary plot, in particular eliminating values of $\omega_A$ above 20\% or below 5\% and setting an upper (lower) limit on the values of $\omega_B$ ($\omega_C$). These observations are quantified in Figure~\ref{fig:Ex2_degeneracy}, in which we plot in the $(\tilde\omega_B, \tilde\omega_C)$ plane all points in our scan with $\Delta\Theta < 0.001$. In the left panel of the figure the points are color-coded by the value of $\tilde\omega_A$. The plot thus outlines the bottom of the dark blue valley in the ternary plot in Figure~\ref{fig:Ex2_degeneracy_ternary}, quantifying the relation among the three concentrations along the degeneracy. In the right panel of Figure~\ref{fig:Ex2_degeneracy} the points are instead color-coded by the ratio
\begin{equation}
    \left(\frac{R_S}{H}\right)_{retrieved}\biggl/\left(\frac{R_S}{H}\right)_{true}
\label{eq:ratioofratios}
\end{equation}
of the retrieved and true values for the quantity $R_S/H$. We see that the degeneracy in the solutions for the chemical abundances results in a range of possible values for the ratio (\ref{eq:ratioofratios}). The upper bound can be understood from the left panel in Figure~\ref{fig:Ex2_degeneracy_ternary}, where there is a $\vec L_\perp$ vector of maximal length along the measured direction of $\vec r_\perp$. Note that the allowed family of vectors $\vec L_\perp (\tilde \omega_a)$ includes the origin of the coordinate system, where $|\vec L_\perp|=0$ (this is why, to avoid division by zero, in this section we chose to calculate $R_S/H$ instead of $H/R_S$). The inclusion of the origin leads to the singularity point which is visible in the ternary plot as the point of convergence of all $\Delta \Theta$ contours, since $\Delta \Theta$ is undefined at the origin. 

This concludes our set of toy examples. Before moving on to a more realistic example in the next section, let us use the left panel in Figure~\ref{fig:Ex2_degeneracy_ternary} to comment on one more interesting scenario. Recall that the number of absorbers $N_{abs}$ is {\em a priori} unknown, and we may easily find ourselves in a situation where we have overlooked a possible contributor. For example, consider the same three bin case of this section, but suppose that we did not include component $C$ in our analysis. The $\vec r_\perp$ calculated from the observed spectrum would still point in the same direction as shown in the left panel of Figure~\ref{fig:Ex2_degeneracy_ternary}, and will be inconsistent with {\em any} combination of $A$ and $B$ alone. This would unambiguously point to the presence of an additional unknown absorber.

\section{Example with WFC3 Binning and Three Absorbers} 
\label{sec:toy13}

In the previous three sections we were considering toy examples with relatively few bins and absorbers, but this was done only for easy visualization and to illustrate the geometrical connection. In fact, our method is applicable for an arbitrarily large number of bins $N_{bin}$ and an arbitrarily large number of absorbers $N_{abs}\le N_{bin}-1$; the method would still produce the correct answers, as derived in Section~\ref{sec:inversion}. In other words, our previous low-dimensional examples provide convenient visualizations, but this is not required for the method to be successful.

\begin{figure*}[t]
\centering
\includegraphics[width=.9\textwidth]{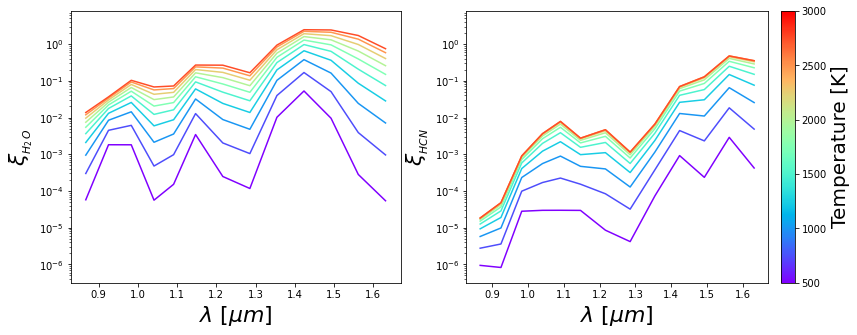}
\caption{Wavelength dependence of the dimensionless absorption coefficients $\xi_a$ defined in (\ref{eq:xidef}) for water (left panel) and HCN (right panel). Each panel shows results for 10 different values of the temperature, starting from $T=500$ K and increasing by $\Delta T= 250$ K.
\label{fig:chi_vs_t}}
\end{figure*}

To demonstrate this, here we shall consider a more complex example motivated by the available data from the Hubble Space Telescope Wide Field Camera 3 (WFC3). Following previous studies in the literature, we shall consider $N_{bin}=13$ bins in the wavelength range $0.838-1.666\, {\mu}m$  \citep{Kreidberg2015ApJ,Marquez2018}. For our study here we shall choose $N_{bin}=3$ absorbers that are typically included in the analysis of transit spectra of hot Jupiters, namely water ($H_2O$), hydrogen cyanide ($HCN$) and gray clouds. The choice of chemical absorbers uniquely fixes the vector arrays $\vec\xi_a$ which are central to any inversion analysis. In our previous examples, we have been assuming constant $\vec\xi$'s, but in reality they depend on the temperature and (to a lesser extent) on the pressure. Therefore, to obtain reliable results from our inversion procedure, we need to account at least for the temperature dependence $\vec\xi_a(T)$. Figure~\ref{fig:chi_vs_t} shows the wavelength dependence of the dimensionless absorption coefficients $\xi_a$ defined in (\ref{eq:xidef}), using the WFC3 binning for water (left panel) and HCN (right panel) in a temperature range starting from $T=500$ K and increasing by $\Delta T= 250$ K. At the same time, for the gray clouds we shall continue to use a constant opacity vector $\vec\xi_0=\vec u$ as prescribed in (\ref{eq:clouds}).

Let us now define a study point with temperature $T=2000$ K and the following parameters for the absorbers
%w_H2O    0.34227 
%w_HCN    0.51341
%w_Clouds 0.14432
\begin{subequations}
\begin{eqnarray}
m_{H_2O} = 18\, m_{amu},~ \quad X_{H_2O}=10^{-2} ~~&\Longrightarrow&~~ \omega_{H_2O} = 0.342;\\
m_{HCN} = 27\, m_{amu}, \quad X_{HCN}=10^{-2} ~~&\Longrightarrow&~~ \omega_{HCN} = 0.513;\\
m_0 = 2.4\, m_{amu}, \quad ~~~X_0 = 10^{-1.5} ~~&\Longrightarrow&~~ \omega_{clouds} = 0.144.
\end{eqnarray}
\label{eq:ex4inputs}
\end{subequations}

\begin{figure*}[t]
\centering
\includegraphics[width=.5\textwidth]{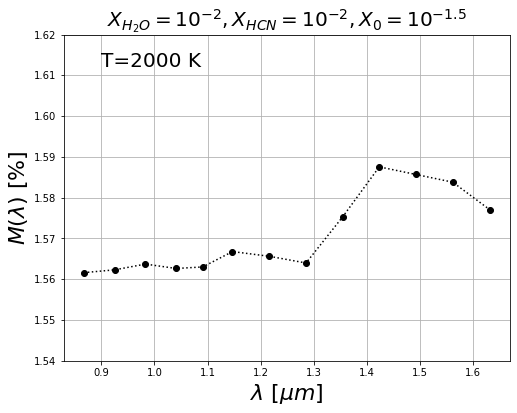}
\caption{The transit spectrum for the study point (\ref{eq:ex4inputs}) analyzed in Section~\ref{sec:toy13}.
\label{fig:M_vs_lambda}}
\end{figure*}

Figure~\ref{fig:M_vs_lambda} shows the generated transit spectrum for our study point (\ref{eq:ex4inputs}). This is the starting point for our retrieval algorithm. 
% T = 2000, X=-2,-2,-1.5
%[1.56162, 1.56231, 1.56372, 1.56264, 1.56302, 1.56679, 1.56563, 1.56399, 1.5753 , 1.58757, 1.5857 , 1.58381, 1.57707]
Next we calculate $\vec r$\, from eq.~(\ref{eq:Mdef}) and decompose it into $\vec r_\parallel$ and $\vec r_\perp$ components, where for the latter, we use a system of $N_{bin}-1=12$ orthonormal basis vectors orthogonal to $\vec u$. 
\begin{figure*}[t]
\centering
\includegraphics[width=.40\textwidth]{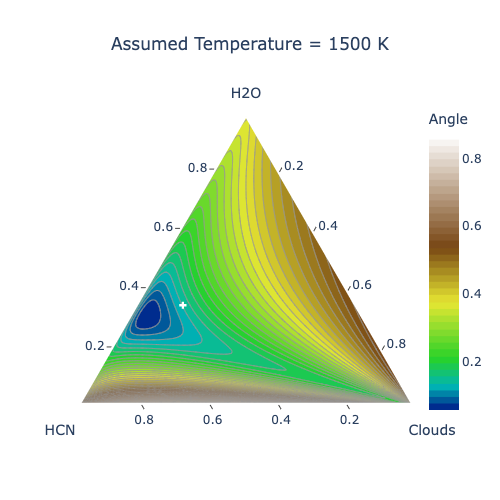}
\includegraphics[width=.40\textwidth]{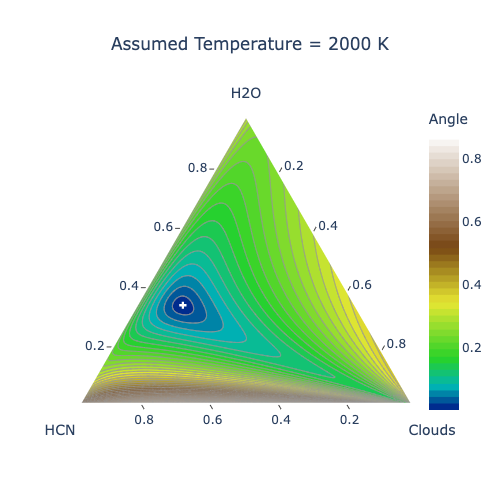}\\
\includegraphics[width=.40\textwidth]{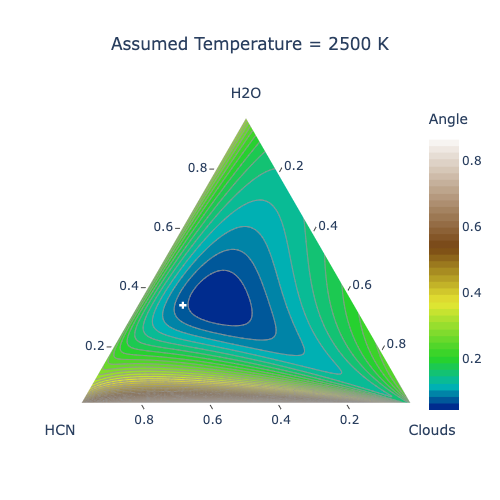}
\includegraphics[width=.40\textwidth]{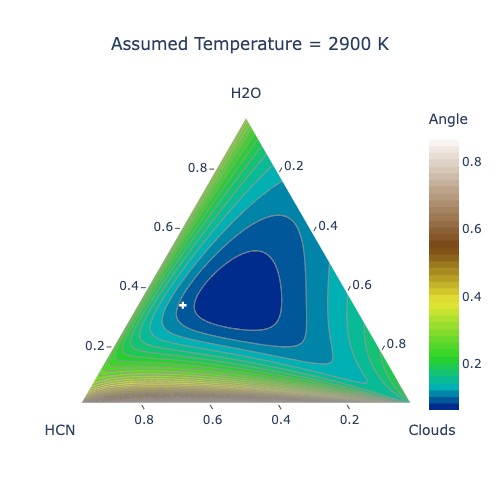}
\caption{Ternary plots analogous to Figure~\ref{fig:TernaryEx2}, which illustrate the extraction of the weights $\omega_A$, $\omega_B$ and $\omega_C$. Each panel corresponds to a different {\em a priori} trial guess $\tilde T$ for the atmospheric temperature, as indicated in the titles. The input temperature for this example was $T=2000$ K. The points are color-coded by the value of the relative angle $\Delta \Theta$ (in radians) between the fixed measured vector $\vec r_\perp$ and the corresponding trial vector $\vec L_\perp (\tilde\omega_A,\tilde\omega_B,\tilde\omega_C,\tilde T)$ from (\ref{eq:LperpsEx3}). As before, the white cross marks the true values of $\omega_A=0.342$, $\omega_B=0.513$ and $\omega_C=0.144$. 
\label{fig:Ex3_2000_trialTemps}}
\end{figure*}
% r_parallel = 0.4518738709293078
% r_perp = whatever
The next step is to construct the grid of vectors 
\begin{equation}
\vec L (\tilde\omega_A,\tilde\omega_B,\tilde\omega_C,\tilde T)\equiv \overrightarrow{\ln \left( \tilde\omega_A\, \vec{\xi}_A(\tilde T) + \tilde\omega_B\, \vec{\xi}_B(\tilde T) + \tilde\omega_C\, \vec{\xi}_C(\tilde T) \right)},
\label{Ex3_Lgrid}
\end{equation}
where for clarity we have explicitly indicated the dependence not only on the trial values $\tilde \omega_a$ for the weights, but also on the trial value $\tilde T$ for the temperature. Next we project (\ref{Ex3_Lgrid}) on the transverse space to obtain the family of transverse vectors
\begin{equation}
\vec L_\perp (\tilde\omega_A,\tilde\omega_B,\tilde\omega_C,\tilde T) 
= \vec L(\tilde\omega_A,\tilde\omega_B,\tilde\omega_C,\tilde T) - (\vec L(\tilde\omega_A,\tilde\omega_B,\tilde\omega_C,\tilde T) \cdot \hat u)\, \hat u .
\label{eq:LperpsEx3}
\end{equation}

The final step is to find which of these $\vec L_\perp$ vectors matches the direction of the measured transverse vector $\vec r_\perp$. This is a trivial computational task and the result is illustrated in the ternary plots  of Figure~\ref{fig:Ex3_2000_trialTemps}, where in analogy to Figure~\ref{fig:TernaryEx2}, we show the relative angle between the two vectors, $\Delta\Theta$, which was introduced in eq.~(\ref{eq:deltathetadef}). Since the absorption coefficients $\xi_a$ are temperature dependent, the retrieval analysis requires a choice for the trial temperature $\tilde T$. Figure~\ref{fig:Ex3_2000_trialTemps} shows results for four such choices of $\tilde T$: $\tilde T=1500$ K (upper left panel), $\tilde T=2000$ K (upper right panel), $\tilde T=2500$ K (lower left panel), and $\tilde T=2900$ K (lower right panel). In interpreting these plots, one should pay attention to two things: i) the location of the found minimum of $\Delta\Theta$ relative to the true answer marked with the white cross, and ii) the value, $\Delta\Theta_{min}$, of the angle $\Delta\Theta$ at the minimum. We see that with the correct guess of the trial temperature ($\tilde T=2000$ K), the minimum is found at precisely the white cross and furthermore, the fit is very good, since $\Delta\Theta_{min}=0.0004$ (for a scan resolution of $\Delta\, \tilde\omega = 0.002$). On the other hand, if we make the wrong guess for the trial temperature, the found minimum shifts, but more importantly, the fit gets worse, since $\Delta\Theta_{min}$ increases. 

\begin{figure*}[t]
\centering
\includegraphics[width=.9\textwidth]{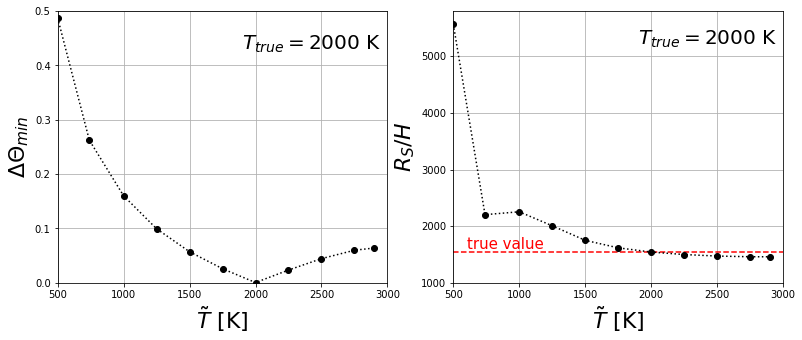}
\caption{The quality of the retrieval procedure as a function of the assumed trial temperature $\tilde T$. The left panel shows the minimum angle $\Delta\Theta$ found in the fit with a scan resolution of $\Delta\, \tilde\omega = 0.002$. An ideal fit with $\Delta\Theta_{min}=0$ is achieved for the true value of the temperature $T=2000$ K. The right panel shows the corresponding retrieved value of $R_S/H$ from (\ref{eq:HRSmeasurement}), where the horizontal red dashed line denotes the true value $R_S/H=1542$. 
\label{fig:theta_vs_t}}
\end{figure*}

The impact of the initial guess $\tilde T$ on the goodness of fit is further investigated in the left panel of Figure~\ref{fig:theta_vs_t}, in which we plot the minimum angle $\Delta\Theta_{min}$ found in the fit as a function of $\tilde T$. As already discussed, the ideal fit, i.e., $\Delta\Theta_{min}=0$, is achieved for the true value of the temperature, $T=2000$ K. As the guess for $\tilde T$ moves away from the true temperature, the fit gets worse, and eventually gives inadequate results at very low temperatures. This behavior of $\Delta\Theta_{min}(\tilde T)$ can in fact be used to {\em measure} the correct temperature $T$ of the atmosphere from
\begin{equation}
T = \argmin_{\tilde T}\left(\Delta\Theta_{min}(\tilde T)\right).
\label{eq:Tmeasurement}
\end{equation}
This idea is similar to the way particle physicists extract the mass $m$ of an invisible particle from the behavior of a {\em transverse} kinematic function $f(\tilde m)$ of a trial mass $\tilde m$ for the invisible particle --- compare the left panel in Figure~\ref{fig:theta_vs_t} to Figure 4 in \cite{Konar:2009wn}.

The right panel of Figure~\ref{fig:theta_vs_t} shows the corresponding retrieved value of $R_S/H$ from eq.~(\ref{eq:HRSmeasurement}) as a function of the trial temperature $\tilde T$. We see that at low temperatures, where the fit for $\Delta\Theta$ is not very good anyway, the extracted value of $R_S/H$ is unreliable. However, in the region around the true temperature $T=2000$ K, the retrieved $R_S/H$ is relatively close to its nominal value marked with the horizontal red dashed line. It is worth noting that at high $\tilde T$ the ratio $R_S/H$ still remains well constrained. 

\section{Summary and Conclusions}
\label{sec:summary}

This section summarizes the main results of the paper and discusses their implications. 

In this paper we proposed a novel method to perform an {\em analytical} inversion of transit spectroscopic observations. The method has a clear geometrical interpretation --- it treats each observed spectrum as a single vector $\vec r$ in the $N_{bin}$-dimensional spectral space. The main idea of the method is to decompose the observed $\vec r$ into i) a wavelength-independent component $\vec r_\parallel$ corresponding to the spectral mean across all observed bins, and ii) a transverse component $\vec r_\perp$ which is wavelength-dependent and contains the relevant information about the atmospheric chemistry. The method allows us to extract, without {\em any} prior assumptions or additional information, the following parameters:
\begin{itemize}
    \item The magnitude of $\vec r_\parallel$, which is equal to the following combination of planet-specific parameters
\begin{equation}
r_\parallel = \biggl\{ \frac{R_0}{R_S} + \frac{H}{R_S} \biggl[\gamma_E + \frac{\vec L\cdot \vec u}{N_{bin}} + \ln \biggl(\mu_{abs}\frac{P_0 \chi }{g}\sqrt{2\pi \frac{R_0}{H}}\biggr) \biggr]\biggr\} \sqrt{N_{bin}}.
\label{eq:rparallelresult}
\end{equation}
    \item The relative mass mixing ratios $\omega_a$ of the absorbers present in the atmosphere. These quantities represent the relative contributions of the different absorbers to the total opacity of the atmosphere and in our method are completely determined by the {\em direction} of the observed $\vec r_\perp$ via eq.~(\ref{eq:abundances}). If desired, these relative mass mixing weights can be easily converted to relative {\em volume mixing} ratios 
\begin{equation}
x_a \equiv \frac{X_a}{\sum_b X_b}    
\end{equation}    
for the absorbers, using the known values $m_a$ for the molecular masses and the measured $\omega_a$ as follows
\begin{equation}
x_a = \frac{\omega_a/m_a}{\sum_b \omega_b/m_b}.   
\label{eq:relativexadef}
\end{equation}
Note that while the {\em absolute} mass or volume mixing ratios are desirable, they suffer from the theoretical degeneracies discussed, e.g., in \cite{Griffith2014,Heng2017,Welbanks2019,matchev2021analytical}. In contrast, the {\em relative} abundances are uniquely determined with our method and that is why we introduced them in our main result (\ref{eq:RTvectorfinal}).
    \item The dimensionless ratio $H/R_S$, which is extracted from the observed {\em magnitude} of $\vec r_\perp$ via eq.~(\ref{eq:HRSmeasurement}). If the stellar radius is known from other independent observations, this yields a robust determination of the scale height $H$ by itself.
    \item The atmospheric temperature $T$, which can be determined via eq.~(\ref{eq:Tmeasurement}). This measurement relies on the sensitivity of the absorption coefficients $\chi_a$ to the temperature of the atmosphere.
\end{itemize}

The main goal of this paper was to establish the theoretical fundamentals of our method, which was then successfully demonstrated in several examples of increasing complexity. In the cases where we had sufficient information, i.e., when $N_{bin}\ge N_{abs}+1$, we were able to correctly reproduce the input values for the above quantities, which validates the method. The further application of the method to real exoplanet transit observations is currently in progress. 

The method has an important built-in cross-check which allows to judge the reliability of the fitted results. For example, using the wrong set of chemical absorbers in the analysis will result in a poor match of the $\vec r_\perp$ direction as measured by the value of $\Delta\Theta_{min}$.

By taking into account the wavelength dependence of the observed transit spectrum, the theoretical analysis presented here completes the theoretical discussion of degeneracies given in \cite{matchev2021analytical}. In particular, all of the different classes of degeneracies discussed in that paper are also manifestly present in eq.~(\ref{eq:rparallelresult}). The new element here is that we are able to determine the relative chemical composition of the atmosphere and its temperature (due to the wavelength and temperature dependence of the absorption coefficients).

\begin{acknowledgments}
This work was supported in part by the United States Department of Energy under Grant No. DESC0022148.
\end{acknowledgments}


\software{
{\tt jupyter} \citep{Kluyver2016},
{\tt matplotlib} \citep{Hunter:2007ouj},
{\tt numpy} \citep{vanderWalt2011},
{\tt plotly} \citep{plotly},
{\tt scipy} \citep{Scipy2020}.
}



%\newpage
\bibliography{analyticalinv}{}
\bibliographystyle{aasjournal}



\end{document}